\documentclass[english]{article}
\usepackage[T1]{fontenc}
\usepackage[latin9]{inputenc}
\usepackage{amsmath}
\usepackage{amssymb}
\usepackage{esint}
\usepackage{babel}
\begin{document}

\title{Constructing Quantum Mechanics from a Clifford substructure of the
relativistic point particle}

\author{Kaare Borchsenius%
\thanks{Bollerisvej 8, 3782 Klemensker, Denmark.%
}}

\date{28 November 2013}
\maketitle
\begin{abstract}
We show that the quantized free relativistic point particle can be
understood as a string in a Clifford space which generates the space-time
coordinates through its inner product. The generating algebra is preserved
by a unitary symmetry which becomes the symmetry of the quantum states.
We start by resolving the space-time canonical variables of the point
particle into inner products of Weyl spinors with components in a
Clifford algebra. Next, we show that a system of $N$ particles has
a $U(N)$ symmetry that mixes the Clifford coordinates and momenta
belonging to different particles. The inner products of these variables
are assembled into Hermitian matrices $X$ and $P$ which are employed
in defining a general unitarily invariant dynamical system. When $X$
and $P$ commute, this system can be gauged back into the original
system of independent particles. When they do not commute, the system
becomes irreducible and infinite and generates a space\nobreakdash-time
canonical system formally identical to Matrix Mechanics. The continuum
limit is identified as a particular parametrization of a relativistic
string in Clifford space.
\end{abstract}

\section{Introduction}

There are two reasons why Clifford algebras are interesting for a
deeper understanding of the relationship between quantum mechanics
and space-time structure. The first one is that in even dimensions
they come with a built\nobreakdash-in unitary symmetry in their generating
algebra which can serve as a basis for the unitary symmetry of quantum
states. This is the subject matter of this paper. The second reason,
as argued in \cite{key-1}, is that quantum mechanics can be understood
as a local theory on such a non\nobreakdash-commutative space. This
resolves the apparent paradox that quantum mechanics, though formulated
as a local theory, shows non\nobreakdash-local behavior. We shall
briefly elaborate on this second point. 

Unlike other statistical theories, quantum mechanics employs probabilities
that are not primary quantities, but are expressed in terms of underlying
linear complex amplitudes. When applied to experiments like the single
particle double slit experiment, this leads to interference terms
in the probabilities which signal an apparent non-local behavior.
This does not show that quantum mechanics \textit{per se} is non-local,
only that it is non-local with respect to space-time (or any equivalent
commutative space). If space-time was generated by an underlying space
with the structure group $SL(2.C)$, it is not difficult to imagine
that the double homomorphism $SL(2.C)\rightrightarrows SO(1.3)$ would
permit a local interpretation of the interference terms. For mathematical
reasons such a space would necessarily have to be non-commutative
and therefore Bell's theorem \cite{key-2} would not apply. In \cite{key-1}
(see also \cite{key-3,key-4,key-5}), we studied a model of this kind.
The space-time coordinates $x^{\mu}$ of the relativistic point particle
were resolved into spinors with components in a Clifford algebra according
to
\[
\sigma_{\mu}^{A\dot{B}}x^{\mu}=c^{A}\bullet c^{*\dot{B}}
\]
where $\sigma$ are the Pauli matrices, $c^{A}$ transforms like a
two-component Weyl spinor and $\bullet$ is the inner product of the
Clifford algebra. The easiest way to understand how this can affect
locality is to consider the transition amplitudes in terms of paths
in Clifford space. In \cite{key-1} we showed that, in a suitable
parametrization, the simplest possible classical trajectories for
the relativistic point particle are

\begin{equation}
c^{A}=a^{A}\tau\qquad x^{\mu}=b^{\mu}\tau^{2}\qquad\tau\in\mathbb{R}\label{eq:a1}
\end{equation}
The trajectories in Clifford space cover the trajectories in space-time
twice with $c(\tau)$ and $c(-\tau)$ corresponding to the same space-time
point $x(\tau^{2})$. In the quantum regime, however, paths are not
restricted in this way and can contain points corresponding to different
positions in space at the same proper time $\tau^{2}$. When in the
double slit experiment a particle travels from one point to another
through two slits, there are only two alternative sets of paths in
space-time but four alternative sets of paths in Clifford space. The
transition amplitude in Clifford space is the sum of four parts and
is identical to the space-time transition probability. The two interference
terms in the transition probability are simply the amplitudes for
the particle to travel along a path in Clifford space which passes
through both slits at opposite values of $\tau$. According to this
view, the resolution of the locality problem follows from the topology
of the Lorentz group and is implemented by a non-commutative representation.

The Clifford model differs in one respect from conventional physics,
in that the world line of a particle cannot be extended into both
the infinite past and the infinite future as measured in proper time.
In space-time, the endpoint of the trajectory would appear as a singular
point, but in Clifford space it merely represents a {}`turning point'
where the space-time trajectory is being reproduced for the second
time.

\section{Mathematical preliminaries}

It is well known that a null vector can be resolved into a product
of two Weyl spinors
\begin{equation}
x^{A\dot{B}}=c^{A}\cdot c^{*\dot{B}}\qquad x^{\mu}x_{\mu}=0\label{eq:b2}
\end{equation}
where $x^{A\dot{B}}$ and $x^{\mu}$ are related through the equivalence
between real four-vectors and second-rank hermitian spinors

\begin{equation}
V^{\mu}=\frac{1}{2}\sigma_{A\dot{B}}^{\mu}V^{A\dot{B}}\qquad V^{A\dot{B}}=\sigma_{\mu}^{A\dot{B}}V^{\mu}\label{eq:b1}
\end{equation}
and $\sigma_{\mu}$ are the four hermitian Pauli matrices. To resolve
non-null vectors, we need something like
\begin{equation}
x^{A\dot{B}}=c^{A}\bullet c^{*\dot{B}}\label{eq:b3}
\end{equation}
where $\bullet$ is a product which belongs to some non-commutative
algebra. This problem can be compared to the somewhat similar problem
of resolving the Lorentz metric $\eta_{\mu\nu}$ into vectors. The
well known solution is $\eta_{\mu\nu}=\frac{1}{2}\{\gamma_{\mu},\gamma_{\nu}\}$
where the Dirac matrices $\gamma_{\mu}$ generate the Clifford algebra
$\mathcal{C}l(1,3,\mathbb{R}$). The components of any real symmetric
$4\times4$ matrix of signature $(1,3)$ can therefore be expressed
as the inner products (anti-commutators) of vectors (real linear combinations
of $\gamma$ matrices) belonging to $\mathcal{C}l(1,3,\mathbb{R})$.
Real Clifford algebras are associated with real quadratic forms, but
there is no similar connection between hermitian sesquilinear forms
and complex Clifford algebras $\mathcal{C}l(\mathbb{C})$ \cite{key-12}.
Instead we must use even-dimensional real Clifford algebras written
in complex form. Consider a future directed time-like vector $x^{\mu}$.
A unitary transformation followed by a non-uniform scaling can reduce
$x^{A\dot{B}}$ to a diagonal matrix with ones in the diagonal and
can be effected by a suitable linear transformation of $c^{A}$ so
that (\ref{eq:b3}) becomes
\begin{equation}
c_{i}\bullet c_{j}^{*}=\delta_{ij}\label{eq:b4}
\end{equation}
This can be compared to the algebra of creation and annihilation operators
for two fermions

\begin{equation}
\{\mathbf{a}_{i},\mathbf{a}_{i}^{\dagger}\}=\delta_{ij}\cdot1\qquad\{\mathbf{a}_{i},\mathbf{a}_{j}\}=0\qquad i,j=1,2\label{eq:b5}
\end{equation}
Defining $e_{i}=\mathbf{\mathit{i}(a}_{i}+\mathbf{a}_{i}^{\dagger})\,,\,\,\, e_{2+i}=\mathbf{a}_{i}-\mathbf{a}_{i}^{\dagger}\,,\,\, i=1,2\:$
, the commutation relations (\ref{eq:b5}) become
\begin{equation}
\{e_{i},e_{j}\}=-2\delta_{ij}\qquad i,j=1,\ldots,4\label{eq:b6}
\end{equation}
which generate the Clifford algebra $\mathcal{C}l(0,4,\mathbb{R})$.
This suggests that a solution to (\ref{eq:b3}) would be to use spinors
with values in the split Clifford algebra $\mathcal{C}l(4,4,\mathbb{R})$
and to let $\bullet$ be the inner product (anti-commutator) of this
algebra. Since any null vector can be created by letting $c$ contain
only a single generator (giving (\ref{eq:b2})), we do not need to
consider degenerate algebras. This expectation is borne out by the
following proposition
\begin{quote}
\emph{Let} $V_{C}$ \emph{be a} $2n$\emph{\nobreakdash-dimensional
complex linear space with complex conjugation} {*} \emph{and} $H$
\emph{an} $n\times n$\emph{ Hermitian matrix of arbitrary signature.
Then the components of} $H$ \emph{can be expressed as}
\[
H_{ij}=c_{i}\bullet c_{j}^{*}\qquad c_{i}\bullet c_{j}=0\qquad i,j=1,\ldots,n
\]

\emph{where} $c_{i}$ \emph{belong to} $V_{C}$ \emph{and} $\bullet$
\emph{is the inner product}
\[
a\bullet b\equiv\frac{1}{2}\{a,b\}
\]

\emph{of the Clifford algebra} $\mathcal{C}l(2n,2n,\mathbb{R})$ \emph{on
the} $4n$ \emph{dimensional real linear space} $V_{R}$ \emph{which
corresponds to }$V_{C}$ .
\end{quote}
Proof. Let $e_{i},\,\, f_{i},\,\, i=1,\ldots,n$ be a basis for $V_{C}$
and $g_{i}=i(e_{i}+e_{i}^{*}),\: g_{n+i}=e_{i}-e_{i}^{*},\: h_{i}=i(f_{i}+f_{i}^{*}),\: h_{n+i}=f_{i}-f_{i}^{*},\: i=1,\ldots n$
a basis for $V_{R}$ . Let $g_{i}$ and $h_{i}$ generate the Clifford
algebra $\mathcal{C}l(2n,2n,\mathbb{R})$ on $V_{R}$ through

\begin{equation}
g_{i}\bullet g_{j}=2\delta_{ij}\qquad h_{i}\bullet h_{j}=-2\delta_{ij}\qquad g_{i}\bullet h_{j}=0\qquad i,j=1,\ldots,2n\label{eq:b9}
\end{equation}
Then the basis $e_{i},\, f_{i}$ for $V_{C}$ satisfies
\begin{equation}
e_{i}\bullet e_{j}^{*}=-\delta_{ij}\qquad f_{i}\bullet f_{j}^{*}=\delta_{ij}\qquad e_{i}\bullet e_{j}=f_{i}\bullet f_{j}=0\qquad i,j=1,\ldots,n\label{eq:b10}
\end{equation}
We can create any $n\times n$ diagonal matrix of plus or minus ones
by setting $c_{i}$ equal to either $f_{i}$ or $e_{i}$ . A zero
in the $k$\nobreakdash-th entry of the diagonal can be created by
$c_{k}=e_{k}+f_{k}$. A non-uniform scaling followed by a unitary
transformation can transform this diagonal matrix into any desired
$n\times n$ hermitian matrix with the same signature and can be effected
by a suitable complex linear transformation of the $c$\textquoteright{}s
$\Box$ 

We shall resolve both the coordinates and momenta of the point particle
into Clifford spinors

\begin{equation}
x^{A\dot{B}}=c^{A}\bullet c^{\dot{*B}}\qquad p_{A\dot{B}}=d_{A}^{*}\bullet d_{\dot{B}}\label{eq:b11}
\end{equation}
but we also need the Clifford algebra to be large enough that the
inner products $c^{A}\bullet d_{B}^{*}$ are algebraically independent
of $x$ and $p$. This can for example be accomplished by enlarging
$\mathcal{C}l(4,4,\mathbb{R})$ to $\mathcal{C}l(8,8,\mathbb{R})$
and then generating $x$ and $p$ by each their own $\mathcal{C}l(4,4,\mathbb{R})$
subalgebra. This makes $c\bullet d$ vanish. The second step is to
choose two Clifford elements $h_{i}$ whose inner products with both
$c$ and $d$ vanish, and to make the substitution 

\begin{equation}
c^{A}\rightarrow c^{A}+A_{i}^{A}h_{i}\qquad d_{A}^{*}\rightarrow d_{A}^{*}+B_{iA}h_{i}^{*}\label{eq:b13}
\end{equation}
This will only change $x$ and $p$ by additive matrices that will
not constrain them, and the two matrices $A$ and $B$ can be adjusted
to produce any desired value of $c\bullet d$. Apart from this requirement,
the dimension of the single-particle Clifford algebra is not of any
importance in this paper.

Note that $c^{*\dot{A}}$ and $d_{A}^{*}$ have the same commutation
properties but transform differently under $SL(2.\mathbb{C})$. The
complex conjugation symbol $*$ can therefore not be omitted, as it
often is, because it specifies the commutation properties of the element
in question. It is tacitly assumed that the inner product of elements
of the same kind vanishes, and this will not be written out explicitly.

The variation of a real function $f$ which depends on $c^{A}$ only
through an inner product can be expressed on the form

\begin{equation}
\delta f=\frac{\partial f}{\partial c^{A}}\bullet\delta c^{A}+\frac{\partial f}{\partial c^{*\dot{B}}}\bullet\delta c^{*\dot{B}}
\end{equation}
which defines the {}`derivative' of $f$ with respect to $c$. This
will serve as a convenient notation.

\section{Clifford substructure of the relativistic point particle}

Let the space-time coordinates and momenta of the relativistic point
particle be resolved into Clifford spinors according to (\ref{eq:b11}).
The equations of motion are obtained from the condition that the re\-para\-metri\-zation
invariant action
\begin{equation}
I=4\sqrt{m}\int\sqrt[4]{\frac{1}{2}\frac{dc^{A}}{d\tau}\bullet\frac{dc^{*\dot{B}}}{d\tau}\,\frac{dc_{A}}{d\tau}\bullet\frac{dc_{\dot{B}}^{*}}{d\tau}}\, d\tau\label{eq:c2}
\end{equation}
is stationary under arbitrary variations of $c(\tau)$. The momenta
conjugate to $c$ are

\begin{equation}
d_{A}^{*}\equiv\frac{\partial L}{\partial\frac{dc^{A}}{d\tau}}=\sqrt{m}\,\bigl(\frac{1}{2}\frac{dc^{E}}{d\tau}\bullet\frac{dc^{*\dot{F}}}{d\tau}\;\frac{dc_{E}}{d\tau}\bullet\frac{dc_{\dot{F}}^{*}}{d\tau}\bigl)^{-\frac{3}{4}}\bigl(\frac{dc_{A}}{d\tau}\bullet\frac{dc_{\dot{B}}^{*}}{d\tau}\bigl)\,\frac{dc^{*\dot{B}}}{d\tau}\label{eq:c3}
\end{equation}
and as expected, the Hamiltonian vanishes. A straightforward calculation
using the four-vector rule

\begin{equation}
V_{A\dot{E}}V^{B\dot{E}}=\frac{1}{2}\delta_{A}^{B}V_{F\dot{E}}V^{F\dot{E}}\label{eq:c30}
\end{equation}
shows that the conjugate momenta $d_{A}^{*}$ satisfy the constraint
\begin{equation}
p^{\mu}p_{\mu}-m^{2}=0\label{eq:c4}
\end{equation}
where $p_{\mu}$ are the space-time momenta defined in (\ref{eq:b11}).
This happens to be the same constraint as would have been obtained
from the space-time action $\int\sqrt{\dot{x}^{2}}\, d\tau$. According
to constrained dynamics, the Hamiltonian is proportional to the constraint

\begin{equation}
H(p,e(\tau))=e(\tau)(p{}^{\mu}p_{\mu}-m^{2})\label{eq:c5}
\end{equation}
where $e(\tau)$ is an einbein. This Hamiltonian can also be obtained
from the Polyakov action

\begin{equation}
I=\int3e(\tau)^{-\frac{1}{3}}\sqrt[3]{\frac{1}{2}\frac{dc^{A}}{d\tau}\bullet\frac{dc^{*\dot{B}}}{d\tau}\,\frac{dc_{A}}{d\tau}\bullet\frac{dc_{\dot{B}}^{*}}{d\tau}}+m^{2}\, e(\tau)\, d\tau\label{eq:c21}
\end{equation}
which recovers (\ref{eq:c2}) when the equations of motion for $e(\tau)$
are substituted back into the action. The momenta conjugate to $c$,
are

\begin{equation}
d_{A}^{*}=e(\tau)^{-\frac{1}{3}}\bigl(\frac{1}{2}\frac{dc^{E}}{d\tau}\bullet\frac{dc^{*\dot{F}}}{d\tau}\,\frac{dc_{E}}{d\tau}\bullet\frac{dc_{\dot{F}}^{*}}{d\tau}\bigr)^{-\frac{2}{3}}\bigl(\frac{dc_{A}}{d\tau}\bullet\frac{dc_{\dot{B}}^{*}}{d\tau}\bigr)\,\frac{dc^{*\dot{B}}}{d\tau}\label{eq:c28}
\end{equation}
which can be used to determine the Hamiltonian density
\begin{equation}
H(c,d)=d_{A}^{*}\bullet\frac{dc^{A}}{d\tau}+c.c.-L\label{eq:c24}
\end{equation}
where $c.c.$ denotes the complex conjugate of the previous term and
$L$ is the Lagrangian in (\ref{eq:c21}). A straightforward calculation
gives

\begin{equation}
d_{A}^{*}\bullet\frac{dc^{A}}{d\tau}+c.c.=4e(\tau)^{-\frac{1}{3}}\sqrt[3]{\frac{1}{2}\frac{dc^{A}}{d\tau}\bullet\frac{dc^{*\dot{B}}}{d\tau}\,\frac{dc_{A}}{d\tau}\bullet\frac{dc_{\dot{B}}^{*}}{d\tau}}
\end{equation}

\begin{equation}
p^{\mu}p_{\mu}\equiv\frac{1}{2}d{}^{A}\bullet d{}^{*\dot{B}}\, d{}_{A}\bullet d{}_{\dot{B}}^{*}=e(\tau)^{-\frac{4}{3}}\sqrt[3]{\frac{1}{2}\frac{dc^{A}}{d\tau}\bullet\frac{dc^{*\dot{B}}}{d\tau}\,\frac{dc_{A}}{d\tau}\bullet\frac{dc_{\dot{B}}^{*}}{d\tau}}\label{eq:c19}
\end{equation}
which, when applied to (\ref{eq:c24}), gives the Hamiltonian (\ref{eq:c5})
of constrained dynamics. Hence the first order (Hamiltonian) form
of the action (\ref{eq:c2}) is 

\begin{equation}
I=\int d_{A}^{*}\bullet\frac{dc^{A}}{d\tau}+c.c.-e(\tau)(p{}^{\mu}p_{\mu}-m^{2})\: d\tau\label{eq:c6}
\end{equation}
This action has a global $SL(2.\mathbb{C})$ and $U(1)$ gauge symmetry
with the conserved Noether charges

\begin{equation}
\mathcal{J_{AB}}\equiv d_{A}^{*}\bullet c_{B}+d_{B}^{*}\bullet c_{A}\quad\jmath\equiv i(d_{A}^{*}\bullet c^{A}-c.c.)\label{eq:c9}
\end{equation}
To obtain the correct space-time equations of motion, it is necessary
to assume (as an initial value condition) that they vanish

\begin{equation}
d_{A}^{*}\bullet c_{B}+d_{B}^{*}\bullet c_{A}=0\label{eq:c10}
\end{equation}

\begin{equation}
d_{A}^{*}\bullet c^{A}-c.c.=0\label{eq:c11}
\end{equation}
Since all skew symmetric second rank tensors are proportional to $\epsilon_{AB}$,
(\ref{eq:c10}) gives

\begin{equation}
d_{A}^{*}\bullet c_{B}=\mu(\tau)\,\epsilon_{AB}\qquad\mu(\tau)\equiv\frac{1}{2}d_{E}^{*}\bullet c^{E}\label{eq:c12}
\end{equation}
with (\ref{eq:c11}) saying that $\mu(\tau)$ is real. We shall refer
to this condition as the {}`Noether condition'. The canonical equations
of motion are obtained by independent variation of $c$ and $d$ 

\begin{equation}
\frac{dc^{A}}{d\tau}=\frac{\partial H}{\partial d_{A}^{*}}=\frac{\partial H}{\partial p_{A\dot{E}}}d_{\dot{E}}\qquad\frac{d\, d_{A}^{*}}{d\tau}=-\frac{\partial H}{\partial c^{A}}=-\frac{\partial H}{\partial x^{A\dot{E}}}c^{*\dot{E}}\,(=0)\label{eq:c7}
\end{equation}
Taking the inner product of these equations with $c^{*\dot{B}}$ and
$d$$_{\dot{B}}$ gives

\begin{equation}
\frac{dx^{A\dot{B}}}{d\tau}=2\frac{\partial H}{\partial p_{A\dot{E}}}\: c^{*\dot{B}}\bullet d_{\dot{E}}\qquad\frac{dp_{A\dot{B}}}{d\tau}=-2\frac{\partial H}{\partial x^{A\dot{E}}}\: c^{*\dot{E}}\bullet d_{\dot{B}}\,(=0)\label{eq:c8}
\end{equation}
which by use of the Noether condition (\ref{eq:c12}) become

\begin{equation}
\frac{dx^{A\dot{B}}}{d\tau}=2\frac{\partial H}{\partial p_{A\dot{B}}}\mu(\tau)\qquad\frac{dp_{A\dot{B}}}{d\tau}=-2\frac{\partial H}{\partial x^{A\dot{B}}}\,\mu(\tau)\,(=0)\label{eq:c13}
\end{equation}
In the parametrization
\begin{equation}
\overline{e}(\overline{\tau})=\frac{1}{2m\mu(\bar{\tau})}\label{eq:c17}
\end{equation}
these equations reduce to the canonical equations of motion

\begin{equation}
\frac{dx^{\mu}}{d\overline{\tau}}=\frac{\partial\mathcal{H}}{\partial p_{\mu}}\qquad\frac{dp_{\mu}}{d\overline{\tau}}=-\frac{\partial\mathcal{H}}{\partial x^{\mu}}\,(=0)\qquad\mathcal{H}(x,p)\equiv\frac{1}{2m}(p^{\mu}p_{\mu}-m^{2})\label{eq:c18}
\end{equation}
for a relativistic point particle with proper time $\overline{\tau}$.
This proper time is not defined at points where $\mu$ vanishes. There
will be just one such point and it represents a {}`turning point'
where the space-time trajectory has an endpoint and the underlying
trajectory in Clifford space starts to reproduce it for the second
time. From (\ref{eq:c7}) and the Hamiltonian constraint (\ref{eq:c4}),
we obtain an explicit expression for $\mu(\tau)$ 

\begin{equation}
\frac{d}{d\tau}\mu(\tau)=\frac{d}{d\tau}(\frac{1}{2}d_{E}^{*}\bullet c^{E})=e(\tau)m^{2}\qquad\mu(\tau)=\int_{\tau_{0}}^{\tau}m^{2}e(t)\, dt\label{eq:c14}
\end{equation}
Hence $\mu(\tau)$ is determined by the mass of the particle and the
{}`turning point' $\tau_{0}$ of its motion.

\section{System of N particles with a U(N) symmetry}

Assuming that the Clifford algebra for the point particle is $Cl(2n,2n,\mathbb{R})$,
we can accommodate $N$ particles in $Cl(2nN,2nN,\mathbb{R})$ in
such a way that all inner products between Clifford coordinates and
momenta belonging to different particles vanish. The generating algebra

\begin{equation}
e_{i}^{p}\bullet e_{j}^{*q}=\delta_{ij}\delta_{pq}\, sign(p)\quad e_{i}^{p}\bullet e_{j}^{q}=0\quad i,j=1,\ldots N\quad p,q=1,\ldots2n\label{eq:d20}
\end{equation}
where $sign(p)$ denotes the sign of $e^{p}\bullet e^{*p}$, is preserved
by the $U(N)$ unitary transformation

\begin{equation}
e_{i}^{p}\rightarrow U_{ih}\, e_{h}^{p}\qquad U_{ih}U_{jh}^{*}=\delta_{ij}\label{eq:d21}
\end{equation}
If we assemble the canonical variables $c_{i}^{A}$ and $d_{iA}^{*},\, i=1\ldots,N$
of the $N$ particles into the ket- and bra-vectors $\stackrel{>}{C^{A}}$
and $\stackrel{<}{D{}_{A}}$ respectively, then the corresponding
space-time coordinates and momenta are elements of the $N\times N$
diagonal matrices
\begin{equation}
X^{A\dot{B}}=\stackrel{>}{C^{A}}\bullet\stackrel{<}{C^{\dot{B}}}\qquad P_{A\dot{B}}=\stackrel{>}{D_{\dot{B}}}\bullet\stackrel{<}{D_{A}}\label{eq:d1}
\end{equation}
which trivially satisfy the commutation relations

\begin{equation}
[X^{\mu},X^{\nu}]=[P_{\mu},P_{\nu}]=[X^{\mu},P_{\nu}]=0\label{eq:d2}
\end{equation}
The equations of motion for this dynamical system can be derived from
the sum 

\begin{equation}
I=\int Tr\,(\frac{d}{d\tau}\stackrel{>}{C^{A}}\bullet\stackrel{<}{D_{A}}+c.c.-H\,)\, d\tau\qquad H\equiv e(\tau)(P^{\mu}P_{\mu}-m^{2}\cdot\underline{1})\label{eq:d6}
\end{equation}
of the single-particle actions (\ref{eq:c6}). The Noether condition
(\ref{eq:c12}) becomes

\begin{equation}
\stackrel{>}{C^{A}}\bullet\stackrel{<}{D_{B}}=\mu(\tau)\,\delta_{B}^{A}\cdot\underline{1}\label{eq:d7}
\end{equation}
We observe that (\ref{eq:d6}) and (\ref{eq:d7}) are preserved by
the global $U(N)$ transformations

\begin{equation}
\stackrel{>}{C^{A}}\rightarrow U\stackrel{>}{\, C^{A}}\qquad\stackrel{<}{D_{A}}\rightarrow\stackrel{<}{D_{A}\,}U^{\dagger}\label{eq:d9}
\end{equation}
which produce the similarity transformations

\begin{equation}
X^{\mu}\rightarrow UX^{\mu}U^{\dagger}\qquad P_{\mu}\rightarrow UP_{\mu}U^{\dagger}
\end{equation}
of the Hermitian matrices $X^{\mu}$ and $P_{\mu}$. Such transformations
create off-diagonal entries in $X$ and $P$ which correspond to artificial
couplings between Clifford coordinates and momenta belonging to different
particles.

The motion of a classical point particle can be described by a set
of integral curves in the phase space $(x^{\mu},p_{\mu})$. From the
foregoing it follows that these integral curves consist of eigenvalues
of the Hermitian matrices $X^{\mu}$ and $P_{\mu}$ which are the
dynamical variables of a unitarily invariant system.

\section{Matrix Mechanics}

We have seen that $N$ independent particles in Clifford space leads
to a unitarily invariant dynamical system. The reverse problem is
to determine under which conditions such a type of system can be gauged
back into a set of independent particles. To address this problem,
we must define a unitarily invariant system which relaxes one or more
assumptions associated with independence. A natural starting point
is to define an action principle for the whole system which is equivalent
to the combined action principles for the particles it contains. To
this end, we observe that to make $N$ single-particle actions stationary
is equivalent to making all time-independent linear combinations of
them stationary. This corresponds to the action

\begin{equation}
I=\int\sum_{i=1}^{N}\phi_{i}L_{i}\, d\tau\quad L_{i}=d_{iA}^{*}\bullet\frac{dc_{i}^{A}}{d\tau}+c.c.-H(p_{i},e(\tau))\label{eq:e1}
\end{equation}
where the coefficients $\phi_{i}$ are arbitrary real constants, and
$L_{i}$ are the single-particle Lagrangians. In section 6, the $\phi$'s
will be given a geometrical interpretation as a dilaton field. When
$\Phi$ denotes the $N\times N$ diagonal matrix with $\phi_{i}$
along its diagonal, the action (\ref{eq:e1}) can be written as
\begin{equation}
I=\int Tr\,\Bigl(\,\Phi(\,\frac{d\stackrel{>}{C^{A}}}{d\tau}\bullet\stackrel{<}{D_{A}}+h.c.-H\,)\Bigr)\, d\tau\qquad H\equiv e(\tau)(P^{\mu}P_{\mu}-m^{2}\cdot\underline{1})\label{eq:e6}
\end{equation}
where $P_{\mu}$ is diagonal. This action is preserved by the unitary
transformation (\ref{eq:d9}) with $\Phi$ transforming according
to

\begin{equation}
\overline{\Phi}\rightarrow U\Phi U^{\dagger}\qquad\Phi^{\dagger}=\Phi\label{eq:e30}
\end{equation}
The diagonal matrices $\Phi$ and $P_{\mu}$ trivially satisfy the
unitarily invariant conditions

\begin{equation}
\frac{d\Phi}{d\tau}=0\label{eq:e0}
\end{equation}

\begin{equation}
[\Phi,P_{\mu}]=0\label{eq:e3}
\end{equation}

\begin{equation}
[P_{\mu},P_{\nu}]=0\label{eq:e4}
\end{equation}
Conversely, these conditions ensure that the action (\ref{eq:e6})
can be gauged back into (\ref{eq:e1}). The conserved $SL(2.\mathbb{C})$
and $U(N)$ Noether charges corresponding to the action (\ref{eq:e6}),
are
\begin{equation}
J_{AB}=Tr\Bigl(\bigl(\Phi\stackrel{>}{(C_{A}}\bullet\stackrel{<}{D_{B}}+\stackrel{>}{C_{B}}\bullet\stackrel{<}{D_{A}})\Bigr)\qquad j=i(\Phi\stackrel{>}{C^{A}}\bullet\stackrel{<}{D_{A}}-h.c.)
\end{equation}
Requiring that they vanish for all values of $\Phi$, gives (\ref{eq:d7}).
The equations of motion are obtained by requiring the action (\ref{eq:e6})
to be stationary for all $\Phi$ which satisfy (\ref{eq:e0}). By
independent variation of $C$ and $D$, we obtain 

\begin{equation}
\frac{d}{d\tau}\stackrel{>}{C^{A}}=\frac{\partial H}{\partial P_{A\dot{E}}}\stackrel{>}{D_{\dot{E}}}\qquad\frac{d}{d\tau}\stackrel{<}{D_{A}}=-\stackrel{<}{C^{\dot{E}}}\frac{\partial H}{\partial X^{A\dot{E}}}\:(=0)\label{eq:d4}
\end{equation}
Taking the inner product on both sides of these equations with $\stackrel{<}{C^{\dot{B}}}$
and $\stackrel{>}{D}_{\dot{B}}$ and applying (\ref{eq:d7}) and the
re\-para\-metri\-zation (\ref{eq:c17}), we obtain

\begin{equation}
\frac{dX^{\mu}}{d\overline{\tau}}=\frac{\partial\mathcal{H}}{\partial P_{\mu}}\qquad\frac{dP_{\mu}}{d\overline{\tau}}=-\frac{\partial\mathcal{H}}{\partial X^{\mu}}\:(=0)\qquad\mathcal{H}\equiv\frac{1}{2m}(P^{\mu}P_{\mu}-m^{2}\cdot\underline{1})\label{eq:d5}
\end{equation}

Equations (\ref{eq:e6})-(\ref{eq:e4}) describe a general class of
unitarily invariant dynamical systems which includes, but is not limited
to, systems of independent particles. Systems of independent particles
are obtained by adding the commutation relations
\begin{equation}
[X^{\mu},X^{\nu}]=[X^{\mu},P_{\nu}]=0\label{eq:e7}
\end{equation}
allowing all off-diagonal entries of $X$ and $P$, that is all couplings
between different particles, to be gauged away in the same unitary
frame. (\ref{eq:e7}) can be generalized by observing that from the
equations of motion (\ref{eq:d5}) and the commutativity (\ref{eq:e4})
of the space-time momenta, it follows that the skew symmetric tensor
\begin{equation}
J^{\mu\nu}\equiv X^{\mu}P^{\nu}-X^{\nu}P^{\mu}
\end{equation}
is a constant of motion. Constraints on $J^{\mu\nu}$ are therefore
compatible with the equations of motion. A natural choice is to let
$J^{\mu\nu}$ be Hermitian and satisfy the $SO(1.3)$ Lie algebra
\begin{equation}
[J^{\mu\nu},J^{\rho\sigma}]=ik(\eta^{\nu\rho}J^{\mu\sigma}-\eta^{\mu\rho}J^{\nu\sigma}-\eta^{\nu\sigma}J^{\mu\rho}+\eta^{\mu\sigma}J^{\nu\rho})\label{eq:d18}
\end{equation}
which is accomplished by the commutation relations
\begin{equation}
[X^{\mu},X^{\nu}]=0\label{eq:e27}
\end{equation}

\begin{equation}
[X^{\mu},P_{\nu}]=ik\delta_{\nu}^{\mu}\cdot\underline{1}\label{eq:e5}
\end{equation}
For $k\neq0$, the couplings between different particles can no longer
be gauged away and we obtain an infinite and irreducible system of
coupled tracks. It should be noted that according to (\ref{eq:e5}),
$X$ diverges when $P$ approaches diagonality. This does not affect
the action (\ref{eq:e6}), it being independent of $X$. The commutation
relations (\ref{eq:e5}) allow the derivatives of $\mathcal{H}$ to
be written as commutators, turning (\ref{eq:d5}) into

\begin{equation}
\frac{dX^{\mu}}{d\overline{\tau}}=\frac{i}{k}[\mathcal{H},X^{\mu}]\qquad\frac{dP_{\mu}}{d\overline{\tau}}=\frac{i}{k}[\mathcal{H},P_{\mu}]\;(=0)\label{eq:e23}
\end{equation}
These equations together with the commutation relations (\ref{eq:e4}),(\ref{eq:e27})
and (\ref{eq:e5}) are formally identical to Matrix Mechanics in the
Heisenberg picture. 

Matrix Mechanics can conveniently be expressed in a {}`picture'-independent
form by means of an auxiliary gauge connection which turns the global
unitary symmetry into a local one and which couples only to the (vanishing)
$U(N)$ Noether charge. This gauge connection transforms according
to

\begin{equation}
\Gamma\rightarrow U\Gamma U^{\dagger}-i\frac{dU}{d\tau}U^{\dagger}\qquad\bar{\Gamma}(\bar{\tau})=\Gamma(\tau)\frac{d\tau}{d\bar{\tau}}\label{eq:d10}
\end{equation}
and defines the gauge covariant derivatives

\begin{equation}
\nabla_{\tau}\stackrel{>}{V}\equiv\Bigl(\frac{d}{d\tau}-i\Gamma(\tau)\Bigl)\stackrel{>}{V}\qquad\bar{\nabla}_{\bar{\tau}}\stackrel{>}{V}\equiv\Bigl(\frac{d}{d\bar{\tau}}-i\bar{\Gamma}(\bar{\tau})\Bigl)\stackrel{>}{V}\label{eq:d8}
\end{equation}
which turn (\ref{eq:e23}) into

\begin{equation}
\overline{\nabla}_{\bar{\tau}}X^{\mu}=\frac{i}{k}[\mathcal{H},X^{\mu}]\qquad\overline{\nabla}_{\bar{\tau}}P_{\mu}=\frac{i}{k}[\mathcal{H},P_{\mu}]\;(=0)\label{eq:e24}
\end{equation}
The \mbox{Heisenberg} picture corresponds to the gauge $\Gamma=0$.
In the gauge $\overline{\Gamma}(\overline{\tau})=-\frac{1}{k}\mathcal{H}$,
the commutators on the left and right hand sides of (\ref{eq:e23})
cancel out and $X$ and $P$ become stationary. This gauge therefore
corresponds to the Schrödinger picture.

Note that in the classical system $k=0$, both the number of tracks
and the initial values of the canonical variables are arbitrary and
have to be put in by hand, whereas in the non-classical system the
canonical commutation relations (\ref{eq:e5}) determine both the
number of tracks (as being infinite) and automatically provide an
infinite range of (stationary) eigenvalues.

\section{The state vector}

The dynamical system constructed in the foregoing generates a set
of tracks which can be used to describe a physical point particle.
Let us first consider the classical system $k=0$. In the gauge where
$X$ is diagonal, all tracks are decoupled from each other and we
expect that when, for example, the space-time position $x$ of the
particle is being measured at some time $\tau$, a good measurement
will return a value $x_{i}(\tau)$ belonging to one of these tracks.
Expressed in a gauge invariant manner, this is equivalent to saying
that it will return an eigenvalue of $X(\tau)$. The result of a measurement
can be represented as a gauge invariant expectation value $E$ in
terms of a state vector $|\, s>$
\begin{equation}
E(\stackrel{>}{C^{A}})\equiv<s\,|\,\stackrel{>}{C^{A}}\qquad E(X^{A\dot{B}})\equiv E(\stackrel{>}{C^{A}})\bullet E(\stackrel{<}{C^{\dot{B}}})=<s|X^{A\dot{B}}|s>\label{eq:d11}
\end{equation}
$\stackrel{>}{C}(\tau)$ can be expanded in terms of the Clifford
coordinates $c_{i}(\tau)$ which generate the eigenvalues of $X$ 

\begin{equation}
\stackrel{>}{C^{A}}(\tau)=\sum_{r}|\, x_{r}(\tau)\rangle c_{r}^{A}(\tau)\qquad c_{r}^{A}(\tau)\bullet c_{s}^{*\dot{B}}(\tau)=\delta_{rs}x_{s}^{A\dot{B}}(\tau)\label{eq:d12}
\end{equation}
where $|\, x_{i}(\tau)\rangle$ denotes the eigenvectors of $X^{\mu}(\tau)$
with eigenvalues $x_{i}^{\mu}(\tau)$. For short, we shall also refer
to $c_{i}(\tau)$ as eigenvalues. It follows that, if the expectation
value $E(\stackrel{>}{C})$ is going to return the correct value $c_{i}$
of a measurement, the state vector $|\, s>$ must be set equal to
the eigenvector $|\, x_{i}>$. Conversely, if the expectation value
coincides with an eigenvalue of $X$, we would expect a good measurement
to return this value. To serve the purpose of predicting the outcome
of future measurements, the state vector must be subject to a time
evolution. In classical dynamics it is natural to assume that after
a measurement has been performed, the expectation value must stay
on the track corresponding to this measurement. According to the foregoing,
the classical system can be gauged into a set of decoupled tracks
with $X(\tau)$ being diagonal and $\Gamma=0$ . In this gauge the
eigenvectors $|\, x_{i}>$ can be chosen to be constants of motion
and the state vector must therefore also be a constant of motion.
This leads to the gauge invariant time evolution

\begin{equation}
\nabla_{\tau}|s>\equiv(\frac{d}{d\tau}-i\Gamma)\,|s>=0\label{eq:d13}
\end{equation}

For the classical system, these measurement principles merely represent
a different way of formulating the traditional initial value problem.
They do, however, also apply to the non-classical system, irrespective
of the fact that the way they were derived is no longer valid. In
the non-classical system where $X$ and $P$ do not commute, the assumption
that measurements must be expressed through a state vector, imposes
restrictions on which type of measurements that can be performed.
The time evolution (\ref{eq:d13}) also holds true, as follows from
the fact that the state vector is known to be a constant of motion
in the \mbox{Heisenberg} picture $\Gamma=0$. To help appreciate
the difference between the classical and the non-classical systems,
we expand the expectation value $E(C)$ in terms of the eigenvalues
$c_{i}$ 

\begin{equation}
E(\stackrel{>}{C^{A}}(\tau))\equiv<s\,|\stackrel{>}{C^{A}}(\tau)=<s\,|\, x_{i}(\tau)>c_{i}(\tau)
\end{equation}
In the classical system, in the gauge $\Gamma=0$, both $<s\,|$ and
$|\, x_{i}>$ are constants of motion and hence the expectation value
$E(C(\tau))$ is equal to one of the eigenvalues $c_{i}(\tau)$. The
outcome of a measurement is therefore predictable. This is not surprising
since it was used to derive the time evolution of the state vector.
In the non-classical system, in the gauge $\Gamma=0$, the state vector
is also stationary, but the eigenvectors $|\, x_{i}>$ undergo a unitary
time evolution. After a measurement has been performed, the expectation
value therefore drifts into a complex linear combination of different
eigenvalues $c_{i}(\tau)$. Accordingly, the outcome of a measurement
is no longer predictable, but instead occurs with statistical frequencies
given by the Born rule.

The classical and non-classical systems are related through Ehrenfest's
theorem. The time evolution of the expectation value $E(C)$ is 

\begin{equation}
\frac{d}{d\tau}E(\stackrel{>}{C^{A}})=(\nabla_{\tau}<s|\,)\stackrel{>}{C^{A}}+<s|\,\nabla_{\tau}\stackrel{>}{C^{A}}=\langle s|\,\frac{i}{2k}[H,X^{A\dot{E}}]\stackrel{>}{D_{\dot{E}}}\label{eq:d14}
\end{equation}
Taking the inner product of this equation with $\stackrel{<}{E(C^{\dot{B}})}$
and using (\ref{eq:d7}) and (\ref{eq:d11}), we obtain after a re\-para\-metri\-zation
the time evolution of the expectation value of the space-time coordinates 

\begin{equation}
\frac{d}{d\overline{\tau}}E(X^{\mu})=\langle s\,|\,\frac{i}{k}[\mathcal{H},X^{\mu}]\,|\, s\rangle\label{eq:d16}
\end{equation}
This is Ehrenfest's theorem which could also have been obtained directly
from (\ref{eq:e24}) by use of (\ref{eq:d11}) and (\ref{eq:d13}). 

In the non-relativistic limit, the proper time $\overline{\tau}$
is equal to the expectation value of $X^{0}$ which represents the
{}`physical' time $t\equiv<s|X^{0}|s>$

\begin{equation}
\frac{dt}{d\overline{\tau}}=<s|\bar{\nabla}_{\overline{\tau}}X^{0}|s>=\frac{1}{m}<s|P^{0}|s>\approx1
\end{equation}
where we have used the time evolution of the state vector and the
equations of motion for $X^{0}$. Restricting the equations of motion
(\ref{eq:e23}) to $\mu=1,2,3\:$, the Hamiltonian $\mathcal{H}$
effectively reduces to the non-relativistic Hamiltonian

\begin{equation}
\tilde{H}=\frac{1}{2m}(P_{x}^{2}+P_{y}^{2}+P_{z}^{2})
\end{equation}
Taken together with the corresponding commutation relations, this
system is identical to that of non-relativistic Matrix Mechanics.
The Schrödinger picture corresponds to the non-relativistic gauge
condition $\overline{\Gamma}(t)=-\frac{1}{k}\tilde{H}$ which turns
the time evolution (\ref{eq:d13}) of the state vector into the matrix
form of the Schrödinger equation.

\section{The relativistic string in Clifford space}

To obtain the continuum limit of the dynamical system constructed
in section 4, we start by mapping the generators $e_{i}^{p}$ of the
generating algebra (\ref{eq:d20}) for $N=\infty$ into the Clifford
elements 

\begin{equation}
f^{p}(\sigma)=\sum_{i=1}^{\infty}g_{i}(\sigma)e_{i}^{p}
\end{equation}
where $g_{i}(\sigma)$ are complex functions of a real parameter $\sigma$.
These functions are chosen so that $f$ satisfies 
\begin{equation}
f^{p}(\sigma)\bullet f^{*q}(\sigma')=\delta_{n}(\sigma-\sigma')\, sign(p)\delta^{pq}\qquad f(\sigma)\bullet f(\sigma')=0\label{eq:f30}
\end{equation}
where $\delta_{n}(\sigma),\, n=1,\ldots$ is a sequence of positive
even functions which converges to the Dirac delta function $\delta(\sigma)$
for $n\rightarrow\infty$. $f^{p}(\sigma)$ can be regarded as a ket-vector
$\stackrel{>}{f^{p}}$ with a continuous index $\sigma$, and correspondingly
$\delta_{n}(\sigma-\sigma')$ can be regarded as a real symmetric
matrix $\delta_{n}$ with continuous indices $\sigma$ and $\sigma'$.
In this notation, the algebra (\ref{eq:f30}) is preserved by the
pseudo-unitary transformation

\begin{equation}
\stackrel{>}{f^{p}}\rightarrow U\stackrel{>}{f^{p}}
\end{equation}
which preserve the metric $\delta_{n}$ 

\begin{equation}
U\delta_{n}U^{\dagger}=\delta_{n}\label{eq:f37}
\end{equation}
In the continuum limit $n\rightarrow\infty$ these pseudo-unitary
transformations become unitary transformations. The Clifford coordinates
$c(\tau,\sigma)$ are defined as integral transforms of $f$

\begin{equation}
c^{A}(\tau,\sigma)\equiv\int a_{p}^{A}(\tau,\sigma,\sigma')f^{p}(\sigma')\, d\sigma'
\end{equation}
and represent a string in Clifford space. The dynamics of this string
will be derived from an action principle which is preserved by arbitrary
re\-para\-metri\-zations $(\tau,\sigma)\rightarrow(\tau',\sigma')$.

It is well known that for a string which resides in space-time, the
Lorentz metric $\eta_{\mu\nu}$ induces a metric on the worldsheet
through the tangent derivatives $\partial_{\alpha}x^{\mu}$. For a
string which resides in Clifford space, we use the complex vectors
\begin{equation}
V_{\alpha}^{\mu}\equiv\sigma_{A\dot{B}}^{\mu}c^{A}\bullet\partial_{\alpha}c^{*\dot{B}}\label{eq:f1}
\end{equation}
which have the real part $\partial_{\alpha}x^{\mu}$. These vectors
induce the hermitian tensor 

\begin{equation}
g_{\alpha\beta}\equiv V_{\alpha}^{\mu}V_{\beta}^{\nu*}\eta_{\mu\nu}\qquad g_{\alpha\beta}^{*}=g_{\beta\alpha}\label{eq:f2}
\end{equation}
on the Clifford worldsheet, which can be decomposed into a real symmetric
tensor $h_{\alpha\beta}$ and a real scalar $\phi$
\begin{equation}
g_{\alpha\beta}=h_{\alpha\beta}+i\phi\sqrt{h}\,\epsilon_{\alpha\beta},\: h_{\alpha\beta}\equiv g_{(\alpha\beta)},\:\phi\equiv-\frac{1}{2}ih^{-\frac{1}{2}}\epsilon^{\alpha\beta}g_{\alpha\beta},\: h\equiv|det(h_{\alpha\beta})|\label{eq:f4}
\end{equation}
The re\-para\-metri\-zation invariant string generalization of
the Polyakov point particle action (\ref{eq:c21}), is 

\begin{equation}
I=\int\bigl(3\sqrt[3]{W^{\mu}W_{\mu}}-m^{2}\bigl)\phi\sqrt{h}\, d\tau d\sigma\qquad W^{\mu}\equiv\frac{1}{2}\sigma_{A\dot{B}}^{\mu}h^{\alpha\beta}\partial_{\alpha}c^{A}\bullet\partial_{\beta}c^{*\dot{B}}\label{eq:f14}
\end{equation}
The geometrical interpretation of this action is obtained from the
equations of motion for the metric $h_{\alpha\beta}$ 

\begin{equation}
(W^{\mu}W_{\mu})^{-\frac{2}{3}}W_{A\dot{B}}\partial_{(\alpha}c^{A}\bullet\partial_{\beta)}c^{*\dot{B}}-\frac{1}{2}(3(W^{\mu}W_{\mu})^{\frac{1}{3}}-m^{2})h_{\alpha\beta}=0
\end{equation}
Contracting this equation with $h^{\alpha\beta}$ gives $(W^{\mu}W_{\mu})^{\frac{1}{3}}=m^{2}$
and thereby

\begin{equation}
I=2m^{2}\int\phi\sqrt{h}\, d\tau d\sigma
\end{equation}
which, apart from a dilaton field $\phi$, is proportional to the
area of the worldsheet.

To write the action (\ref{eq:f14}) in an explicit covariant first
order form, we use Dedonder-Weyl covariant canonical variables \cite{key-11}.
The multi-momenta conjugate to $c$ are

\begin{equation}
\textrm{d}_{A}^{^{*}\alpha}\equiv\frac{\partial L}{\partial(\partial_{\alpha}c^{A})}=(W^{\nu}W_{\nu})^{-\frac{2}{3}}W_{\mu}\sigma_{A\dot{B}}^{\mu}h^{\alpha\beta}\partial_{\beta}c^{*\dot{B}}\phi\sqrt{h}\label{eq:f6}
\end{equation}
where $L$ denotes the Lagrangian density in (\ref{eq:f14}). With
a redefinition $\textrm{d}\rightarrow\phi\sqrt{h}\, d$ of the momenta,
this leads to the expressions 
\begin{equation}
\frac{1}{2}h_{\alpha\beta}\, d^{*\alpha A}\bullet d^{\dot{\beta B}}\, h_{\gamma\delta}d_{A}^{*\gamma}\bullet d_{\dot{B}}^{\delta}=\sqrt[3]{W^{\mu}W_{\mu}}\label{eq:f7}
\end{equation}

\begin{equation}
d_{A}^{*\alpha}\bullet\partial_{\alpha}c^{A}+c.c.=4\sqrt[3]{W^{\mu}W_{\mu}}\label{eq:f8}
\end{equation}
from which we obtain the Dedonder-Weyl covariant Hamiltonian density

\begin{equation}
H\equiv\phi\sqrt{h}\, d_{A}^{*\alpha}\bullet\partial_{\alpha}c^{A}+c.c.-L=(p^{\mu}p_{\mu}+m^{2})\phi\sqrt{h}\label{eq:f9}
\end{equation}

\begin{equation}
p^{\mu}\equiv\frac{1}{2}\sigma_{A\dot{B}}^{\mu}h_{\alpha\beta}\, d^{*\alpha A}\bullet d^{\beta\dot{B}}\label{eq:f10}
\end{equation}
and hence the first order form

\begin{equation}
I=\int\Bigr(d_{A}^{*\alpha}\bullet\partial_{\alpha}c^{A}+c.c.-(p^{\mu}p_{\mu}+m^{2})\Bigl)\phi\sqrt{h}\, d\tau d\sigma\label{eq:f40}
\end{equation}
of the Polyakov action (\ref{eq:f14}). By independent variation of
$d$ and $c$, we obtain the equations of motion 

\begin{equation}
\partial_{\alpha}c^{A}=h_{\alpha\beta}\, p^{A\dot{E}}d_{\dot{E}}^{\beta}\label{eq:f51}
\end{equation}

\begin{equation}
\partial_{\alpha}(\phi\sqrt{h}\, d_{A}^{*\alpha})=0\label{eq:f52}
\end{equation}
To turn the dilaton into a dynamical field, we add to the Lagrangian
the Weyl invariant term $\kappa\sqrt{h}\, h^{\alpha\beta}\partial_{\alpha}\phi\,\partial_{\beta}\phi$.
The equations of motion for $h_{\alpha\beta}$ then become

\begin{equation}
\frac{\partial\mathcal{L}}{\partial h_{\alpha\beta}}=-2\phi\sqrt{h}\, p^{\mu}\frac{1}{2}\sigma_{\mu}^{A\dot{B}}d_{A}^{*(\alpha}\bullet d_{\dot{B}}^{\beta)}-\kappa\sqrt{h}\, h^{\alpha\gamma}h^{\beta\delta}\partial_{\gamma}\phi\,\partial_{\delta}\phi+\frac{1}{2}\mathcal{L}h^{\alpha\beta}=0\label{eq:f38}
\end{equation}

\begin{equation}
\mathcal{L}\equiv\Bigl(d_{A}^{*\alpha}\bullet\partial_{\alpha}c^{A}+c.c.-(p^{\mu}p_{\mu}+m^{2})\Bigr)\phi\sqrt{h}\,+\kappa\sqrt{h}\, h^{\alpha\beta}\partial_{\alpha}\phi\,\partial_{\beta}\phi\label{eq:f39}
\end{equation}
Upon inserting the expression (\ref{eq:f51}) for $\partial c$ into
(\ref{eq:f39}), the trace of (\ref{eq:f38}) gives the mass shell
equation

\begin{equation}
p^{\mu}p_{\mu}-m^{2}=0\label{eq:f23}
\end{equation}
Note that the kinetic term for $\phi$ does not contribute to the
trace because it is Weyl invariant. (\ref{eq:f38}) determines the
metric algebraically. The metric could be turned into a dynamical
field by adding the term $\phi\sqrt{h}\, h^{\alpha\beta}R_{\alpha\beta}$
to the Lagrangian in accordance with the Jackiw-Teitelboim 2d model
of gravity \cite{key-13,key-14}. However, in the second order formulation
it would break Weyl invariance and violate the mass shell equation
(\ref{eq:f23}), and in the first order formulation, where the affinity
is varied independently of the metric, $\phi$ would become a constant
\cite{key-15}. Varying the action with respect to $\phi$ gives the
equations of motion 

\begin{equation}
2\kappa\frac{1}{\sqrt{h}}\partial_{\alpha}(\sqrt{h}\, h^{\alpha\beta}\partial_{\beta}\phi)=d_{A}^{*\alpha}\bullet\partial_{\alpha}c^{A}+c.c.-(p^{\mu}p_{\mu}+m^{2})\label{eq:f60}
\end{equation}
When (\ref{eq:f51}) and (\ref{eq:f23}) are applied to the right
hand side of (\ref{eq:f60}), it becomes $m^{2}$.

The equations of motion (\ref{eq:f52}) determine the multi-momenta
only up to an arbitrary worldsheet scalar $\chi_{A}$ 

\begin{equation}
\phi\sqrt{h}\, d_{A}^{*\alpha}\rightarrow\phi\sqrt{h}\, d_{A}^{*\alpha}+\epsilon^{\alpha\beta}\partial_{\beta}\chi_{A}^{*}
\end{equation}
The class of solutions which correspond to the discrete model in section
4 is obtained by choosing $\chi_{A}$ so that $d_{A}^{*\beta}$ satisfies
the re\-para\-metri\-zation invariant condition

\begin{equation}
\mu^{\alpha}\epsilon_{\alpha\beta}d_{A}^{*\beta}=0\label{eq:f15}
\end{equation}

\begin{equation}
\mu^{\alpha}\equiv\frac{1}{4}(d_{A}^{*\alpha}\bullet c^{A}+c.c.)\label{eq:f55}
\end{equation}
This condition will be imposed as a constraint on the action principle
by adding the terms $\lambda^{A}\bullet\mu^{\alpha}\epsilon_{\alpha\beta}d_{A}^{*\beta}+c.c.$
to the Lagrangian, where $\lambda^{A}$ is a Lagrange multiplier.
The equations of motion (\ref{eq:f51}) and (\ref{eq:f52}) then become

\begin{equation}
\phi\sqrt{h}\,\partial_{\alpha}c^{A}=\phi\sqrt{h}\, h_{\alpha\beta}\, p^{A\dot{E}}d_{\dot{E}}^{\beta}-\frac{1}{2}c^{A}(\lambda^{E}\bullet\epsilon_{\alpha\beta}d_{E}^{*\beta})-\lambda^{A}\epsilon_{\beta\alpha}\mu^{\beta}\label{eq:f80}
\end{equation}

\begin{equation}
\partial_{\alpha}(\phi\sqrt{h}\, d_{A}^{*\alpha})=-\frac{1}{2}d_{A}^{*\beta}(\lambda^{E}\bullet\epsilon_{\beta\gamma}d_{E}^{*\gamma})\label{eq:f81}
\end{equation}
In the discrete system, the vanishing of the Noether charges was imposed
as an initial value condition. Correspondingly, the conserved $SL(2.\mathbb{C})$
and $U(1)$ Noether currents
\begin{equation}
\mathcal{J_{AB}^{\alpha}}\equiv\phi\sqrt{h}(d_{A}^{*\alpha}\bullet c_{B}+d_{B}^{*\alpha}\bullet c_{A})\qquad\jmath^{\alpha}\equiv i\phi\sqrt{h}(d_{A}^{*\alpha}\bullet c^{A}-c.c.)\label{eq:f16-1}
\end{equation}
are assumed to vanish on some space-like curve on the worldsheet.
In a para\-metri\-zation where $\mu^{2}=0$, it follows from the
constraint (\ref{eq:f15}) that $d^{2}=0$ and therefore also $\mathcal{J}^{2}=\jmath^{2}=0$.
In such a para\-metri\-zation, the Noether charges $\mathcal{J}^{1}$
and $\jmath^{1}$ become constants of motion and must vanish everywhere.
Hence the Noether currents $\mathcal{J}^{\alpha}$ and $\jmath^{\alpha}$
vanish everywhere, leading to the Noether condition

\begin{equation}
d_{A}^{*\alpha}\bullet c^{B}=\mu^{\alpha}\delta_{A}^{B}\label{eq:f17}
\end{equation}
where $\mu^{\alpha}$ is the real vector (\ref{eq:f55}). Taking the
inner product of both sides of (\ref{eq:f80}) with $c^{*\dot{B}}$
and contracting with $\mu^{\alpha}$, the $\lambda$-terms vanish
and it reduces to

\begin{equation}
\mu^{\alpha}\partial_{\alpha}x^{\mu}=2(h_{\alpha\beta}\mu^{\alpha}\mu^{\beta})\, p^{\mu}\label{eq:f22}
\end{equation}
Consider the vector
\begin{equation}
v^{\gamma}\equiv(h_{\alpha\beta}\mu^{\alpha}\mu^{\beta})^{-1}\mu^{\gamma}
\end{equation}
In a domain where $\mu^{\alpha}$ is regular, $v^{2}$ and thereby
$\mu^{2}$ can be made to vanish through a re\-para\-metri\-zation.
Any subsequent re\-para\-metri\-zation of the form $\tau\rightarrow\bar{\tau}(\tau,\sigma)$,
$\sigma\rightarrow\overline{\sigma}(\sigma)$ preserves $v^{2}=0$
and can, since the weight of $v$ is different from 1, be used to
make $v^{1}=2m$. In this para\-metri\-zation, the equations of
motion (\ref{eq:f22}) become

\begin{equation}
\frac{\partial x^{\mu}}{\partial\tau}=\frac{1}{m}\, p^{\mu}
\end{equation}
and (\ref{eq:f81}) reduces to

\begin{equation}
\frac{\partial}{\partial\tau}(\phi\sqrt{h}\, d_{A}^{*1})=0\label{eq:f42}
\end{equation}
From (\ref{eq:f42}) and the mass shell equation (\ref{eq:f23}),
it follows that $\phi^{2}h\, h_{11}^{-1}$ and consequently $\sqrt{h_{11}}\, d_{\dot{A}}^{1}$
and $p^{\mu}$ are constants of motion. Accordingly, the constraint
(\ref{eq:f15}) leads to the re\-para\-metri\-zation invariant
equations of motion for $x^{\mu}$ and $p_{\mu}$
\begin{equation}
\nu^{\alpha}\partial_{\alpha}\, x^{\mu}=2p^{\mu}\qquad\nu^{\alpha}\partial_{\alpha}\, p_{\mu}=0
\end{equation}

We are now in a position to compare the discrete system based on the
action (\ref{eq:e1}) with the continuous system based on the action
(\ref{eq:f40}) subject to the constraint (\ref{eq:f15}). The equations
of motion and constraints corresponding to the action (\ref{eq:e1}),
are

\begin{equation}
\frac{dc_{i}^{A}}{d\tau}=e(\tau)\, p_{i}^{A\dot{E}}d_{i\dot{E}}\qquad\frac{d\, d_{i\dot{A}}}{d\tau}=0\qquad p_{i}^{A\dot{B}}\equiv d_{i}^{*A}\bullet d_{i}^{\dot{B}}
\end{equation}
\begin{equation}
p_{i}^{\mu}p_{i\mu}-m^{2}=0\qquad d_{iA}^{*}\bullet c_{i}^{B}=\mu(\tau)\delta_{A}^{B}
\end{equation}
with no summation over $i$. For the continuous system, we shall use
the para\-metri\-zation $v^{2}=\partial_{\sigma}v^{1}=0$ which
allows for an arbitrary time-re\-para\-metri\-zation $\tau\rightarrow\tau'(\tau)$.
We resolve the metric into zweibeins $h_{\alpha\beta}=\eta_{ab}e_{\alpha}^{a}e_{\beta}^{b}$
and choose local zweibein frames in which $e_{1}^{2}=0$. Defining
$\tilde{d}_{\dot{E}}\equiv e_{\alpha}^{1}\, d_{\dot{E}}^{\alpha}$
which is a scalar under time-re\-para\-metri\-zation, the continuous
system can be written as

\begin{equation}
\frac{\partial c^{A}}{\partial\tau}=e_{1}^{1}\, p^{A\dot{E}}\tilde{d}_{\dot{E}}\qquad\frac{\partial\tilde{d}_{\dot{A}}}{\partial\tau}=0\qquad p^{A\dot{B}}\equiv\tilde{d}^{*A}\bullet\tilde{d}^{*\dot{B}}\label{eq:f70}
\end{equation}

\begin{equation}
p^{\mu}p_{\mu}-m^{2}=0\qquad\tilde{d}_{A}^{*}\bullet c^{B}=e_{1}^{1}\,\mu^{1}\delta_{A}^{B}\label{eq:f71}
\end{equation}
From the para\-metri\-zation condition $\partial_{\sigma}v^{1}=0$
it follows that $h_{11}\mu^{1}$ ($=(e_{1}^{1})^{2}\mu^{1}$) is a
function of $\tau$ only, so that the two systems lead to the same
equations of motion for $x^{\mu}$ and $p_{\mu}$, in accordance with
our previous finding.

\section{Clifford strings and the ontology of quantum mechanics}

Interpreting quantum mechanics \cite{key-7} is made more difficult
by the mathematical leap between the space-time description of classical
and of quantum objects. In the Clifford space description, both the
classical and the quantized point particle are understood as strings
and obey the same three measurement principles: \renewcommand{\labelenumi}{(\roman{enumi})}
\begin{enumerate}
\item \textit{the result of a measurement is an eigenvalue of the observable}
\item \textit{the results of measurements performed at the same parameter-time
can be expressed as expectation values corresponding to a single state
vector}
\item \textit{the state vector is gauge covariantly constant in time (Schrödinger
equation)}
\end{enumerate}
These principles are derived from the classical string but also apply
to the quantum string where they have the well-known {}`unexpected'
consequences. The classical string is reducible and can (in a suitable
gauge) be described as a set of tracks in Clifford phase space, the
points of which generate eigenvalues of both  $X$ and $P$. These
tracks are independent of each other in the sense that they are integral
curves corresponding to the same equations of motion. The role of
the state vector is hereby reduced to the trivial one of selecting
an integral curve. This makes it possible to describe not only the
outcome, but also the object of the measurement as a point particle.
The general case of the irreducible Clifford string however, shows,
that such a picture is misleading. The object of a measurement is
a Clifford string, not a point particle. The point particle is invoked
only to describe the outcome of a measurement.

\section{Conclusion}

We have shown that the quantized free relativistic point particle
can be understood as a particular para\-metri\-zation of a relativistic
string in Clifford space. In obtaining this result, we considered
only the dynamical degrees of freedom corresponding to the constraint
(\ref{eq:f15}). 

There are good reasons to believe that a four-dimensional Minkowski
space does not suffice to accommodate the particle physics of the
Standard Model. The \mbox{Clifford} model discussed in the foregoing
is limited to a four-dimensional Minkowski space because it is based
on complex Weyl spinors. Since Weyl spinors are an integral part of
the model, it is difficult to see how the dimension of space-time
can be increased without replacing the complex numbers with a higher
dimensional Algebra. The complex numbers correspond to the \mbox{Clifford}
algebra $\mathcal{C}l(0,1,\mathbb{R})$. Increasing the dimension,
we find $\mathcal{C}l(0,2,\mathbb{R})$ which corresponds to the quaternions
and $\mathcal{C}l(0,3,\mathbb{R})$ which can be deformed into the
octonions. For algebraic reasons \cite{key-8,key-9,key-10}, such
spinors would be expected to generate a six-dimensional and a ten-dimensional
Minkowski space respectively.


\begin{thebibliography}{References}
\bibitem{key-1}K. Borchsenius: Degenerate space-time paths and the
non-locality of quantum mechanics in a Clifford substructure of space-time,
Math. Phys. EJ, Vol 6, No.4, (2000)

\bibitem{key-2}J. S. Bell: On the Einstein Podolsky Rosen Paradox,
Physics 1, 3, 195\textendash{}200 (1964)

\bibitem{key-3}K. Borchsenius: Clifford spinors and the relativistic
point particle, arXiv:hep-th/0111262 (2001)

\bibitem{key-4}K. Borchsenius: Fermi substructure of space-time,
Int. J. Theor. Phys. 34 (1995) 1863-1870

\bibitem{key-5}K. Borchsenius: Spin substructure of space-time, Gen.Rel.Grav.
19 (1987) 643-648

\bibitem{key-12}I. Porteous: Clifford Algebras and the Classical
Groups, Cambridge Univ. Press, Cambridge (1995)

\bibitem{key-11}H. A. Kastrup: Canonical theories of Lagrangian dynamical
systems in physics, Physics Reports, Volume 101, Issue 1-2 (1983)
1-167.

\bibitem{key-13}C. Teitelboim: Gravitation and Hamiltonian Structure
in Two Space-Time Dimensions, Phys.Lett. B126 , 41 (1983)

\bibitem{key-14}R. Jackiw, in Quantum Theory Of Gravity, p. 403-420,
S.Christensen (ed.) (Adam Hilgar, Bristol, 1983)

\bibitem{key-15}S. Deser: Inequivalence of First and Second Order
Formulations in D = 2 Gravity Models, Found. Phys. 26, 617 (1996)

\bibitem{key-7}Franck Laloë: Do We Really Understand Quantum Mechanics?,
Cambridge University Press (2012)

\bibitem{key-8}John C. Baez: The octonions, Bull. Amer. Math. Soc.
(N.S.) 39 (2002), no. 2, 145\textendash{}205. MR

\bibitem{key-9}C. A. Manogue and J. Schray: Finite Lorentz Transformations,
Automorphisms, and Division Algebras , J. Math. Phys. 34 (1993), 3746\textendash{}3767.

\bibitem{key-10}Tevian Dray and Corinne A. Manogue: Octonionic Cayley
Spinors and E6. arXiv:0911.2255 (2009) \end{thebibliography}
\end{document}